\documentclass{ifae.mod}

\usepackage{graphicx}
\usepackage{multicol} 
\usepackage{color}
\usepackage{amsfonts}
\definecolor{rosso}{cmyk}{0,1,1,0.4}
\definecolor{rossos}{cmyk}{0,1,1,0.55}
\definecolor{rossoc}{cmyk}{0,0.5,1,0.2}
\definecolor{blu}{cmyk}{1,1,0,0.3}
\definecolor{blus}{cmyk}{1,1,0,0.6}
\definecolor{blucc}{cmyk}{1,0.4,0.2,0}
\definecolor{viola}{cmyk}{0,1,0,0.6}
\definecolor{viola2}{cmyk}{0,1,0.2,0.6}
\definecolor{verde}{cmyk}{0.92,0,0.59,0.25}
\definecolor{verdec}{cmyk}{0.92,0,0.59,0.15}
\definecolor{verdes}{cmyk}{0.92,0,0.59,0.5}

\font\tenrsfs=rsfs10 at 12pt
\font\sevenrsfs=rsfs7
\font\fiversfs=rsfs5
\newfam\rsfsfam
\textfont\rsfsfam=\tenrsfs
\scriptfont\rsfsfam=\sevenrsfs
\scriptscriptfont\rsfsfam=\fiversfs
\def\mathscr#1{{\fam\rsfsfam\relax#1}}

\newcommand{\fig}[1]{~\ref{fig:#1}}
\newcommand{\eq}[1]{~{\rm (\ref{eq:#1})}}
\newcommand{\sys}[1]{~{\rm (\ref{sys:#1})}}

\def\circa#1{\,\raise.3ex\hbox{$#1$\kern-.75em\lower1ex\hbox{$\sim$}}\,}

\newcommand{\NP}{Nucl. Phys.}
\newcommand{\PRL}{Phys. Rev. Lett.}
\newcommand{\PL}{Phys. Lett.}
\newcommand{\PR}{Phys. Rev.}
\newcommand{\beq}{\begin{equation}}
\newcommand{\eeq}{\end{equation}}
\newcommand{\diag}{\hbox{diag}\,}

\newcommand{\Ve}{V_{e}}
\newcommand{\Vm}{V_{\mu}}
\newcommand{\Vt}{V_{\tau}}

\def\circa#1{\,\raise.3ex\hbox{$#1$\kern-.75em\lower1ex\hbox{$\sim$}}\,}
\makeatletter

\newcommand{\MeV}{\,\hbox{\rm MeV}}
\newcommand{\eV}{\,\hbox{\rm eV}}

\newcommand{\gcm}{\,\mathrm{g}\,\mathrm{cm}^{-3}}
\newcommand{\erg}{\,\mathrm{erg}}
\renewcommand{\sec}{\,\mathrm{sec}}

\newcommand{\nue}{\nu_e}
\newcommand{\nueb}{\bar\nu_e}
\newcommand{\num}{\nu_\mu}
\newcommand{\numb}{\bar\nu_\mu}
\newcommand{\nut}{\nu_\tau}
\newcommand{\nutb}{\bar\nu_\tau}

\newcommand{\nus}{\nu_s}
\newcommand{\nusb}{\bar\nu_{\rm s}}
\newcommand{\nB}{n_{B}}
\newcommand{\GF}{G_{\rm F}}
\newcommand{\km}{\,\mathrm{km}}
\newcommand{\Feo}{\Phi^0_{\nueb}}
\newcommand{\Fmuo}{\Phi^0_{\numb}}
\newcommand{\Ftauo}{\Phi^0_{\nutb}}
\newcommand{\Fso}{\Phi^0_{\nusb}}
\newcommand{\Fe}{\Phi_{\nueb}}
\newcommand{\Fmu}{\Phi_{\numb}}
\newcommand{\Ftau}{\Phi_{\nutb}}
\newcommand{\Fs}{\Phi_{\nusb}}

\def\art{\@ifnextchar[{\eart}{\oart}}
\def\eart[#1]#2#3#4#5#6{{\rm #2}, {\em #3 \rm #4} {\rm (#6) #5} ({\em #1})}
\def\hepart[#1]#2{{\rm #2, \em#1}}
\newcommand{\oart}[5]{{\rm #1}, {\em #2 \rm #3} {\rm (#5) #4}}

\newcounter{alphaequation}[equation]
\def\thealphaequation{\theequation\hbox to
0.6em{\hfil\alph{alphaequation}\hfil}}
\def\eqnsystem#1{
\def\@eqnnum{{\rm (\thealphaequation)}}
\def\@@eqncr{\let\@tempa\relax \ifcase\@eqcnt \def\@tempa{& & &} \or
  \def\@tempa{& &}\or \def\@tempa{&}\fi\@tempa
  \if@eqnsw\@eqnnum\refstepcounter{alphaequation}\fi
\global\@eqnswtrue\global\@eqcnt=0\cr}
\refstepcounter{equation} \let\@currentlabel\theequation \def\@tempb{#1}
\ifx\@tempb\empty\else\label{#1}\fi
\refstepcounter{alphaequation}
\let\@currentlabel\thealphaequation
\global\@eqnswtrue\global\@eqcnt=0 \tabskip\@centering\let\\=\@eqncr
$$\halign to \displaywidth\bgroup \@eqnsel\hskip\@centering
$\displaystyle\tabskip\z@{##}$&\global\@eqcnt\@ne
\hskip2\arraycolsep\hfil${##}$\hfil& \global\@eqcnt\tw@\hskip2\arraycolsep
$\displaystyle\tabskip\z@{##}$\hfil
\tabskip\@centering&\llap{##}\tabskip\z@\cr}
\def\endeqnsystem{\@@eqncr\egroup$$\global\@ignoretrue} \makeatother

\newcommand{\comment}[1]{}

\begin{document}

\begin{header}
  \title{Sterile Neutrinos\\[2mm] in astrophysical and cosmological sauce\Acknow{\star}}

  \begin{Authlist}
    Marco Cirelli\Iref{yale}

  \Affiliation{yale}{Physics Dept. - Yale University, New Haven, CT 06520, USA}
  \Acknowfoot{\star}{Based on the Proceedings for the 10$^{\rm th}$ International Symposium on Particles, Strings and Cosmology (PASCOS '04), 16-22 August 2004, Northeastern University, Boston, MA, USA and for the XVI Incontri sulla Fisica delle Alte Energie (IFAE), 14-16 April 2004, Torino, Italy.}
  \end{Authlist}

  
  \begin{abstract}
The study of sterile neutrinos has recently acquired a different flavor: being now excluded as the dominant solution for the solar or atmospheric conversions, sterile neutrinos, still attractive for many other reasons, have thus become even more elusive.
The present relevant questions are: which subdominant role can they have? Where (and how) can they show up?
Cosmology and supernov\ae\ turn out to be powerful tools to address these issues. 
With the most general mixing scenarios in mind, I present the analysis of many possible effects on BBN, CMB, LSS, and in SN physics due to sterile neutrinos.
I discuss the computational techniques, present the state-of-the-art bounds, identify the still allowed regions and study some of the most promising future probes. I show how the region of the LSND sterile neutrino is excluded by the constraints of standard cosmology.
  \end{abstract}
  
\end{header}

\vspace{-0.5cm}
\section{Introduction}
The study of sterile neutrinos (namely: additional light fermionic particles that are neutral under all Standard Model gauge forces, but can be a non-negligible ingredient of our world through their mixing with $e, \mu, \tau$ neutrinos) has recently acquired a different flavor: the established solar and atmospheric anomalies seem produced by oscillations among the three active neutrinos~\cite{sun nus, atmo nus} so that their explanation in terms of oscillation into a $\nus$ state (which was viable and fairly popular up to a few years ago) is believed to be now ruled out as the dominant mechanism. This means that the {\bf relevant questions concerning sterile neutrinos nowadays} have become:
$\bullet$which is the {\it subdominant} role still possible for sterile neutrinos in solar and atmospheric neutrinos?
$\bullet$where can we detect the effects of a sterile neutrino? i.e. which are the most sensitive experiments (in astrophysics, cosmology or man-made set-ups) in which sterile neutrinos can be discovered?
$\bullet$how can we detect the effects of a sterile neutrino? i.e. which are the signatures of its presence?

To answer these questions, the investigation on sterile neutrinos requires a more extensive and deep approach.\\
Indeed, for instance, most of the previous analysis used to consider only the peculiar oscillation pattern that gives the simplest physics: the initial active neutrino $|\nu_a \rangle$ ($\nue$ in the case of solar neutrinos, $\num$ in the case of the atmospheric ones) oscillates into an energy-independent mixed neutrino $\cos\theta_s |\nu'_a \rangle + \sin\theta_s |\nu_s \rangle$.
This of course leaves unexplored the largest part of the parameter space.
Sometimes, moreover, previous analysis used to neglect for simplicity the mixing among active neutrinos (this is the case of most studies on the sterile effects in cosmology or in supernov\ae). Such a mixing is now established and its parameters are reasonably pinned down.


These considerations motivate the analysis performed in~\cite{CMSV} and presented here, which  considers and includes:
\begin{itemize}
\item[{\bf a.}] any possible $\nu_{e,\mu,\tau}-\nus$ mixing pattern; 
\item[{\bf b.}] the established $\nue-\nu_{\mu,\tau}$, $\num-\nut$ mixings;
\item[{\bf c.}] all possible neutrino sources and contexts (cosmology (BigBang Nucleosynthesis-BBN, Cosmic Microwave Background-CMB, Large Scale Structures-LSS), astrophysics (the Sun, supernov\ae-SNe...), atmospheric neutrinos, reactor and accelerator experiments...),
\end{itemize}
The purpose is to set the state-of-the-art bounds on the active-sterile mixing parameters (in each context separately and then in a combined way) and to identify the most promising future probes.
Since the different cosmological quantities and the observables of SN physics turn out to be very important to this aim, investigating complementary regions and allowing the application of techniques that extend to other fields, these are the contexts on which I focus in these Proceedings.

\comment{
The main {\bf purposes} of such a study are evident:
\begin{enumerate}
\item we {\bf look for evidences} of sterile neutrino effects in present data from all contexts.\\ 
I anticipate we do find {\bf no evidence} with any statistical significance (the LSND result is addressed separately). Therefore...
\item we {\bf set the present bounds} on the active-sterile mixing parameters in each context separately, and then in a combined way.\footnote{We do not perform a global fit of all data (cosmological, astrophysical, solar, reactor, atmospheric...) together because of the different reliability of the different probes, the diversity of the assumptions and so on. Not to mention the computational effort that such a procedure would require.  
The plot with superimposed bounds presented in the end is in our opinion a better way to present the combined analysis.}
\item we {\bf compare the LSND hint} against those bounds.
\item we {\bf identify the most promising future probes} of sterile neutrino effects in all contexts.
\end{enumerate}
}
\comment{
Before going to the details, a crucial issue that concerns the {\bf motivations} of such a study is still to be addressed. 
Now that the need of sterile neutrinos for the solar and atmospheric anomalies is died, can sterile neutrinos still be interesting ingredient of our world at all?
The answer is yes, for at least two different sets of reasons.\\
}
Before going to the details, let's stress that now that the hunger of sterile neutrinos for the solar and atmospheric anomalies is over, nevertheless {\bf sterile neutrinos are even more attractive} for at least two sets of reasons.\\
First, from a top-down point of view, sterile fermions that are naturally light or cleverly lightened arise in many theories that try to figure out what is beyond the Standard Model. To begin, slightly beyond the context of the SM, the right handed neutrino is an obvious candidate, to complete the lepton sector in similarity (symmetry?) with the quark one. In this case, actually, three states (one per family) would be natural. More broadly, several (SuSy-/GUT-/string-/ED- inspired) SM gauge singlets line up awaiting for consideration (axino, branino, dilatino, familino, Goldstino, Majorino, mirror fermion, modulino, radino...)~\cite{sterile models}. The origin and the load of information of any of these particles can be very different, but from an effective point of view we simply need to parameterize their mixing with the active neutrinos in terms of  mixing angles and $\Delta m^2$ (see below) to include them all in the analysis. Independently from the specific model, the discovery of a new light particle would be of fundamental importance and deserves to be investigated per se.\\
Second, from a more phenomenological bottom-up perspective, sterile neutrinos are repeatedly pointed as a possible explanation for several puzzling situations in particle physics, astronomy and cosmology. For instance, they have been invoked~\cite{sterile puzzles} to account for the origin of the pulsar kicks, to constitute a Dark Matter candidate, to explain (via their decay) the diffuse ionization of the Milky Way, to help the r-process nucleosynthesis in the environment of exploding stars, to interpret the slightly too low Argon production in the Homestake experiment... and notoriously to explain the LSND claimed evidence of oscillations~\cite{LSND}. Any one of these puzzles, in general, points to specific sterile neutrinos (i.e. with a specific mixing pattern with the active ones) and it is therefore worthwhile to explore them in an extensive way.

\comment{
\medskip

In Sec.\ref{4nu} I present the non-standard four neutrino formalism that we use. In Sec.\ref{cosmology} I address the sterile effects in cosmology, comparing the relative sensitivities of
two BBN probes (the $^4$He and Deuterium abundances), of CMB
and of LSS. In Sec.\ref{LSND section} I single out the LSND hint for a sterile neutrino and I discuss the issue of its compatibility with the cosmological constraints.
In Sec.\ref{SN} I study sterile oscillations in SN1987A and future supernov\ae.
The study of the sterile effects in all other contexts (solar, atmospheric neutrinos; reactor, accelerator SBL, LBL neutrinos; high energy astrophysical neutrinos) can be found in~\cite{CMSV}; in Sec.\ref{conclusions} I collect all the combined bounds and draw some conclusions.
}

\section{Four neutrino mix}
\label{4nu}
In absence of sterile neutrinos, we denote by
$U$  the usual $3\times 3$ mixing matrix that relates
neutrino flavor eigenstates $\nu_{e,\mu,\tau}$ to active
neutrino mass eigenstates $\nu_{1,2,3}$
as $\nu_\ell = U_{\ell i} \nu_i$.
The extra sterile neutrino can then mix with a {\bf mixing angle $\theta_s$} with an {\bf arbitrary combination of active neutrinos}, which we identify by a complex unit 3-vector $\vec{n}$
\footnote{In this way $\vec{n}$ and $\theta_s$, together with the four neutrino masses, simply reorganize in a more intuitive way all the parameters of the most generic $4\times 4$ Majorana neutrino mass matrix, which  contains 4 masses, 6 mixing angles and 6 CP-violating phases, of which 3 affect oscillations.
The $4\times 4$ neutrino mixing matrix $V$ that relates flavor to mass eigenstates as $\nu_{e,\mu,\tau,s} = V\cdot \nu_{1,2,3,4}$ is expressed in this parameterization by
\begin{equation}
V = \pmatrix{1-(1-\cos\theta_{\rm s}) \vec{n}^* \otimes \vec{n} & \sin\theta_{\rm s} \vec{n}^* \cr
-\sin\theta_{\rm s} \vec{n}   & \cos\theta_{\rm s}} \times \pmatrix{U&0\cr 0 & 1}
\end{equation}
Commonly, $V$ is instead parameterized as
$ V= R_{34}R_{24}R_{14}\cdot U_{23}U_{13} U_{12}$
when studying sterile mixing with a flavor eigenstate.
Here $R_{ij}$ represents a rotation in the $ij$ plane by angle $\theta_{ij}$
and $U_{ij}$ a complex rotation in the $ij$ plane.
$\theta_{14}$ or $U_{e4}$ give rise to $\nu_e/\nu_{\rm s}$ mixing,
 $\theta_{24}$ or $U_{\mu 4}$ to $\nu_\mu/\nu_{\rm s}$ mixing,
and  $\theta_{34}$ or $U_{\tau 4}$ to $\nu_\tau/\nu_{\rm s}$ mixing.
When studying the mixing with a mass eigenstate, the parameterization of $V$ is changed to 
$ V = U_{23}U_{13}U_{12}\cdot  R_{34}R_{24}R_{14}$. 
Now $\theta_{i4}$ gives rise to $\nu_i/\nu_{\rm s}$ mixing and so on. 
Our parameterization is more convenient because $\vec{n}$ already encrypts in a natural way the information on which active states mix with the $\nus$, while $\theta_{\rm s}$ simply expresses the size of the mixing.
}
\begin{equation}
\vec{n}\cdot \vec{\nu} = n_e \nu_e + n_\mu \nu_\mu + n_\tau \nu_\tau = 
n_1\nu_1 + n_2 \nu_2 + n_3 \nu_3\qquad (n_i = U_{\ell i}n_\ell).
\end{equation}
So, in particular, the 4$^{\rm th}$ mass eigenstate is given by $\nu_4 =  \nu_{\rm s} ~\cos\theta_{\rm s} + n_\ell \nu_\ell ~\sin\theta_{\rm s}$.
We allow all possible values of its {\bf mass $m_4$} (`the mass of the sterile neutrino', in the small mixing case).

\begin{figure}
$$\includegraphics[width=0.4\textwidth]{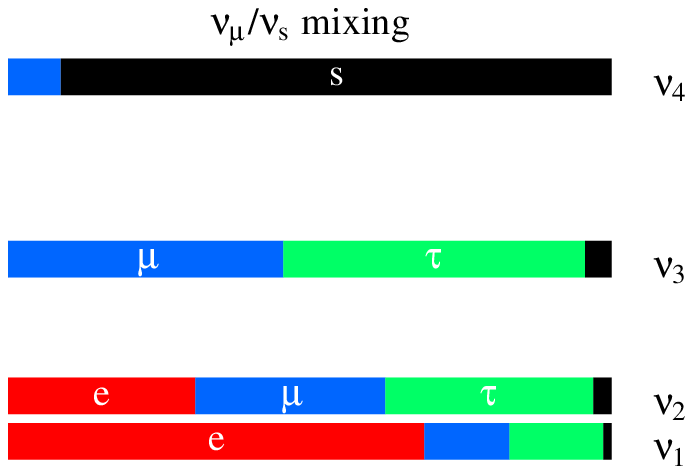}\qquad
\includegraphics[width=0.4\textwidth]{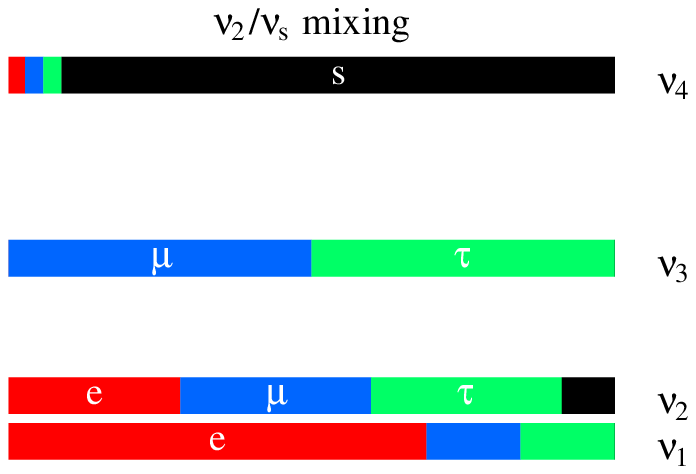}$$
\caption{\label{fig:spettrias}
Basic kinds of four neutrino mass spectra. 
\em Left: sterile mixing with a flavor eigenstate ($\nu_\mu$ in the picture).
Right: sterile mixing with a mass eigenstate ($\nu_2$ in the picture).
}
\end{figure}

Such a formalism is completely general and covers of course all the possible mixing patterns. We need to choose, however, some intuitive limiting cases to present the results in the following:
\begin{itemize}
\item  {\bf Mixing with a flavor eigenstate}  (fig.\fig{spettrias}a):
the sterile neutrino oscillates into one of the active flavors ($\vec{n}\cdot\vec{\nu}=\nu_\ell$ with $\ell =e$ or $\mu$ or $\tau$). Therefore there are 3 different active-sterile $\Delta m^2$ (which cannot all be smaller than the observed splittings $\Delta m^2_{\rm sun,atm}$, see figure).

\item {\bf Mixing with a mass eigenstate}  (fig.\fig{spettrias}b):
the sterile neutrino oscillates into a matter eigenstate, that consists therefore of mixed flavor ($\vec{n}\cdot\vec{\nu}=\nu_i$, with $i=1$ or 2 or 3). There is one single $\Delta m^2$, which can be arbitrarily small.
\end{itemize}

In the case of mixing with a matter eigenstate, we will consider also the situation in which the (mostly) sterile state is lighter than the (mostly) active state with which it mixes (imagine the $\nu_4$ state lowered below $\nu_2$ in figure\fig{spettrias}b): it is represented by the portion of $\theta_{\rm s}>\pi/4$ in our plots.

We assume that active neutrinos have normal hierarchy, $\Delta m^2_{23}>0$.
Finally, we assume $\theta_{13}=0$. We verified that using $\theta_{13}\sim 0.2$,
the maximal value allowed by present experiments,
leads to minor (in some cases) or no (in other cases) modifications,
that we do not discuss. 

\section{Sterile Neutrinos in cosmological sauce}
\label{cosmology}

{\bf Generalities:} The Early Universe can be a powerful laboratory for neutrino physics, and therefore in particular for the physics of sterile neutrinos. The fundamental reasons for this basic fact are simply listed:
\begin{itemize}
\item (light) neutrinos are very abundant (namely ``as abundant as photons'') for a long period of the evolution of the Universe, keeping thermal equilibrium until $T\sim$ few$\ \MeV$;
\item in a Friedman-Robertson-Walker standard cosmology, the total energy density is a crucial parameter that sets the expansion rate of the Universe; since that energy is predominantly stored in the relativistic species, namely electrons, positrons, photons and all species of neutrinos, for $T \simeq 100 \MeV \rightarrow 1 \MeV$, it is evident that the relative abundance of neutrinos (e.g. increased by the presence of additional states) is a very relevant quantity;
\item the early plasma is so dense that neutrinos are initially trapped and undergo peculiar matter effects  while the density decreases as a consequence of the expansion;
\item the detailed balance of the different species of neutrinos among themselves can also be important for processes that distinguish flavor: for instance, the $\nue$ density affects the $n \rightarrow p$ conversion and therefore is imprinted in the primordial ratio of $n/p$ that we read today (see below).  
\end{itemize}

We have access to several different windows during the history of the Universe. 
From them we get a sensitivity to neutrino properties (masses, oscillation parameters...)  that is nowadays competitive with direct measurements and even offers a brighter prospective of improvements in the near future, making the study of quantitative neutrino cosmology worthwhile.

{\bf BBN:} BigBang Nucleosynthesis occurs at $T \sim (1 \div 0.1)\MeV$ and describes the era when the light elements were synthesized~\cite{BBN in general}. Given a few input parameters 
(the effective number $N_\nu$ of thermalized relativistic species,
the baryon asymmetry $n_B/n_\gamma=\eta$, 
and possibly the $\nu_\ell /\bar\nu_\ell$ lepton asymmetries)
BBN successfully predicts the abundances of several light nuclei. 
Today $\eta$ is best determined within minimal cosmology by CMB data to be $\eta = (6.15 \pm 0.25) 10^{-10}$~\cite{WMAP}. 
Thus, neglecting the lepton asymmetries (see below), basically one uses the observations of primordial abundances to test if $N_\nu=3$ as predicted by the SM\footnote{For the sake of precision, relaxing the hypothesis of instantaneous neutrino freeze-out and carefully including the partial neutrino reheating from $e^+e^-$ annihilations, the related spectral distortions and finite temperature QED small effects, the SM prediction is actually $N_\nu \simeq 3.04$~\cite{mangano 3.04}. The deviations due to any exotic phenomenon (including sterile neutrinos) go on top of this.}. This is what is often done~\cite{NnuBBN}, using state-of-the-art, publicly available codes~\cite{BBN code}. 
We need however to do something slightly more refined: indeed, $N_\nu$ is an effective parameter that sums up and does not distinguish the relative contributions of the various neutrinos (3 active and 1 (or more) sterile), which can instead be different according to the mixing patterns; moreover, in general, the neutrino populations can have a peculiar behavior in time (temperature) which would be concealed by the use of $N_\nu$.

For these reasons, in our computation we use as variables the four neutrino densities~\footnote{The densities are intended as relative to the photon one, so that $\rho_{\nu_l} \in (0,1)$.} $\rho_{\nu_e}$, $\rho_{\nu_\mu}$, $\rho_{\nu_\tau}$, $\rho_{\nu_s}$: we follow their evolution with the temperature during the whole period of BBN, for each possible choice of the sterile mixing parameters.
Starting (conservatively) from a zero initial abundance of the sterile neutrino, at a certain point oscillations start producing it~\cite{bbnosc}. This essentially happens when the plasma effects/thermal masses for the (active) neutrinos cease to be dominant compared to the vacuum masses, as the Universe expands and cools; at what point precisely (and how efficiently) the production occurs is something which is determined by the sterile mixing parameters. In the meanwhile, other cosmological processes occur, as standard: at $T\sim {\rm few} \MeV$ neutrinos freeze-out (i.e. loose thermal equilibrium with the bath, but they still take part in the $n \leftrightarrow p$ reactions, see below); at $T\sim 1 \MeV$ $e^+e^-$ annihilate etc...

{\footnotesize

More precisely, we follow the time evolution of the full $4 \times 4$ density matrices $\varrho$ and $\bar{\varrho}$ (written in the flavor basis), of which the four neutrino densities above constitute the diagonal. In absence of neutrino asymmetries, the neutrino and antineutrino sectors decouple and proceed identically, so we focus on the neutrinos for definiteness. The kinetic equations for such matrices must take into account (i) the vacuum oscillations (active-active and active-sterile), (ii) the matter effects in the primordial plasma, (iii) the $\nu e \leftrightarrow \nu e$ scattering reactions and the $\nu \nu \leftrightarrow ee$ annihilation reactions. 
They read~\cite{bbnosc,OscEU,DolgovReview}
\begin{equation}
\frac{d\varrho}{dt} \equiv \frac{dT}{dt} \frac{d\varrho}{dT} = -i \left[ {\mathcal H}_m,\varrho \right] - \left\{\Gamma, (\varrho-\varrho^{\textrm{eq}}) \right\}
\label{eq:kin eq 1}
\end{equation}
${\mathcal H}_m$ is the Hamiltonian in matter, composed by the vacuum Hamiltonian in the flavor basis and the matter potentials $V_l$ for each flavor: these consist of the thermal masses for the neutrinos in the primordial plasma, rapidly decreasing with $T$. The usual MSW potential is in this case subdominant because the plasma is charge symmetric.
\begin{equation}
{\mathcal H}_m =\frac{1}{2 E_\nu}\left[ V \, {\rm diag}(m_1^2,m_2^2,m_3^2,m_4^2) \, V^\dagger  + E_\nu \textrm{diag}(\Ve,\Vm,\Vt,0) \right]  
\end{equation}

\begin{equation}
\label{eu potentials}
\begin{array}{rcl}
V_e &=& -\frac{199 \sqrt{2}\pi^2}{180}\frac{\zeta(4)}{\zeta(3)}\GF \frac{T_\nu}{M_W^2}\left(T^4+\frac{1}{2}T_\nu^4\cos\theta_W\ \varrho_{ee} \right) \\
V_{\mu,\tau} &= & -\frac{199 \sqrt{2}\pi^2}{180}\frac{\zeta(4)}{\zeta(3)}\GF \frac{T}{M_W^2}\left(\frac{1}{2}T_\nu^4\cos\theta_W\ \varrho_{\mu\mu,\tau\tau} \right) \\
V_s &= &0
\end{array}
\end{equation}

Concerning the reaction part, it can be shown that one has to use: in the equations for the off-diagonal components of $\varrho$, $\Gamma_{\rm tot}  \approx  3.6 \; G_{\rm F}^2 \; T^5$ for $\nu_e$ and $\Gamma_{\rm tot}  \approx  2.5 \; G_{\rm F}^2 \; T^5$ for $\nu_{\mu,\tau}$; in the equations for the diagonal components, $\Gamma_{\rm ann} \approx 0.5 \; G_{\rm F}^2 \; T^5$ for $\nu_e$ and $\Gamma_{\rm ann}\approx  0.3 \; G_{\rm F}^2 \; T^5$ for $\nu_{\mu,\tau}$.  $\varrho^{\rm eq} = {\rm diag}(1,1,1,0)$ is the equilibrium value of the density matrix to which the reactions tend. The neutrino freeze-out occurs when these $\Gamma$'s are overwhelmed by the expansion of the Universe.

The determination of $\frac{dT}{dt}$ is in principle quite involved, since we need to keep track of the several phenomena that go on in the range $T\sim 1 \MeV$ which is under examination. In particular, we want to include the possible extra degrees of freedom (the sterile neutrinos) that are produced by the oscillations and we do not want to neglect the heating due to $e^+ e^-$ annihilations. In first approximation, we can use the standard expression $\dot{T}=-H(\rho_{\nu_{\rm tot}})\ T$, where $H$ contains the (temperature dependent) total energy density, including in particular that of all neutrinos.   


After solving the neutrino densities evolution with temperature, we can study the relative neutrons/protons abundance, which is the all-important quantity for the outcome of primordial nucleosynthesis, since those are the building blocks of the nuclei that are going to be formed and essentially all neutrons are incorporated into some light element in the process: the neutron abundance at the moment that the synthesis begins practically fixes the proportions of all the products. $n/p$ evolves according to
\begin{equation}
\dot{r} \equiv \frac{dT}{dt} \frac{dr}{dT} = \Gamma_{p \to n} (1-r)- r \Gamma_{n \to p} \qquad
r = \frac{n_n}{n_n+n_p}.
\end{equation}
$\Gamma_{p \to n}$ is the total rate for all the $p \to n$ reactions ($n \to p\ e^{-}\ \nueb,
n\ \nue \to p\ e^-, n\ e^+ \to p\ \nueb$) and $\Gamma_{n \to p}$ for the inverse processes. They depend on the $\rho_{\nu_e}$ and $\rho_{\bar\nu_e}$ densities computed above. 
At this point, it is apparent how the production of {\bf sterile neutrinos can enter the game} by 
\begin{itemize}
\item[{\bf A.}] entering in $\rho_{\nu_{\rm tot}}$ and thus increasing the Hubble parameter $H$ i.e. the expansion rate;
\item[{\bf B.}] modifying the $\Gamma_{p\to n},\Gamma_{n\to p}$ rates directly,  if the $\nue,\nueb$ population is depleted by oscillations.
\end{itemize}

With the value of $n/p$ in hand, finally a network of Boltzmann equations describes how
electroweak, strong and electromagnetic processes
control the evolution of the various nuclei: $p$, $n$, D, T, $^3$He, $^4$He,\ldots

For the purposes of the CMB and LSS bounds discussed below, the neutrino densities 
at the time of recombination and today are needed. These are simply given by the final outputs (i.e. for $T \ll 0.1\MeV$) of the kinetic equations.
}

We assume a vanishing or negligibly small {\bf neutrino asymmetry} $\eta_\nu$. Allowing an unnatural large $\eta_\nu$ (namely, much larger than the corresponding asymmetry $\eta$ in baryons),  besides affecting non trivially the evolution~\cite{asymm}, means adding an extra relevant parameter, so that one could essentially conceal any sterile effect, at least as long as observations will not be able to break the degeneracy.

\begin{figure}[t]
\begin{center}
\includegraphics[width=0.590\textwidth]{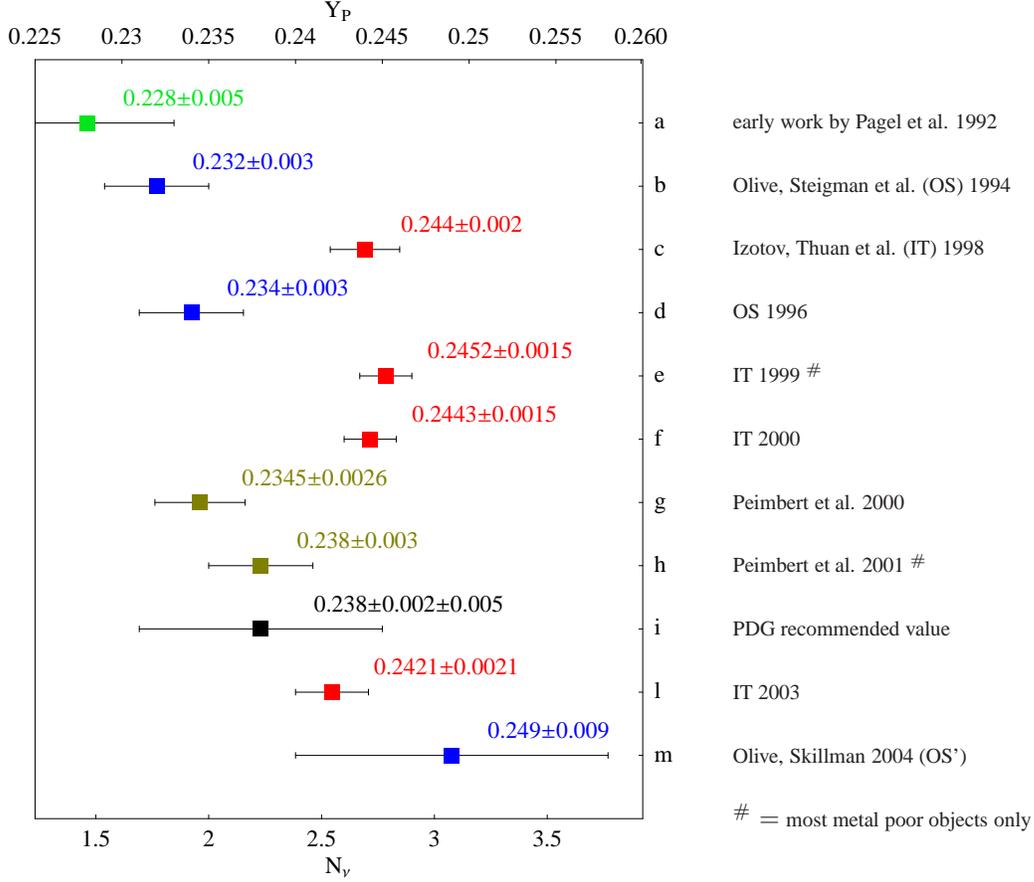}
\hspace{0.1cm}
\raisebox{5.38cm}{\parbox[b]{5cm}{  
\begin{tabular}{l}
{\footnotesize early work by Pagel et al. 1992}\\
\\
{\footnotesize Olive, Steigman et al. (OS) 1994}\\
\\
{\footnotesize Izotov, Thuan et al. (IT) 1998}\\
\\
{\footnotesize OS 1996}\\
\\
{\footnotesize IT 1999 $^{\#}$}\\
\\
{\footnotesize IT 2000}\\
\\
{\footnotesize Peimbert et al. 2000}\\
\\
{\footnotesize Peimbert et al. 2001 $^{\#}$}\\
\\
{\footnotesize PDG recommended value}\\
\\
{\footnotesize IT 2003}\\
\\
{\footnotesize Olive, Skillman 2004 (OS')}\\
\\
$^{\#} = $ \footnotesize{most metal poor objects only} 
\end{tabular}
}}
\end{center}
\caption{Some recent experimental results in the determination of the primordial $^4$He abundance $Y_p$. \em The bars correspond to the claimed $1\sigma$ errors. The PDG recommended value adds an estimate of the systematic uncertainties ($\pm 0.005$). References are in~\cite{He determinations}.
}
\label{fig:He}
\end{figure}

At the end of the process, we can compute how the light elements abundances are modified and compare them to the observations.
We focus on the $^4$He abundance, which is today the most sensitive probe of sterile effects, but we also study the Deuterium abundance, which has brighter prospects of future improvements. The observational determinations of both quantities are plagued by controversial systematic uncertainties.
For instance, Fig.\ref{fig:He} collects some of the most recent results for $^4$He~\cite{He determinations}. A similar situation, with perhaps less controversy and more overall uncertainty, applies to the case of Deuterium~\cite{D determinations}.
Conservative estimates are
\beq\begin{array}{l}\label{eq:pD}
 Y_p = 0.24\pm 0.01,  \\
 \displaystyle
 Y_{\rm D}/Y_{\rm H} =  (2.8 \pm 0.5)\,10^{-5} ,
\end{array}\eeq
where $Y_X  \equiv n_{X}/n_B$ and $Y_p$ is the traditional notation for $Y_{^4{\rm He}}$.

At this point, for ease of presentation, we can even convert the values of the computed primordial abundances back into effective numbers of neutrinos, $N_\nu^{^4{\rm He}}$ and $N_\nu^{\rm D}$.
For arbitrary values around the SM value of 3, BBN codes predict~\cite{BBN in general,NnuBBN} the following relations
\begin{eqnsystem}{sys:HeD}
\label{eq:BBNfit}
Y_p &\simeq& 0.248+0.0096\ln\frac{\eta}{6.15~10^{-10}} + 0.013 (N_\nu^{^4{\rm He}}-3),\\
Y_{\rm D}/Y_{\rm H} &\simeq &(2.75\pm 0.13)~10^{-5}~\frac{1+0.11~(N_\nu^{\rm D}-3)}{(\eta/6.15~10^{-10})^{1.6}}. 
\end{eqnsystem}
The observed abundances of eq.(\ref{eq:pD}) then translate into
\beq
\begin{array}{c}\label{eq:NpD}
 N_\nu^{^4{\rm He}} \simeq  2.4\pm0.7,    \\
 N_\nu^{\rm D}\simeq 3\pm 2.
\end{array}
\eeq

\medskip

{\bf LSS:} Neutrinos can also be studied looking at the distribution of galaxies. The connection lies at the time of the formation of the anisotropies in the primordial plasma that were the seeds for the formation of the Large Scale Structures (which took place much time later).
The point is that the neutrinos, relativistically traveling through the plasma (from which they were decoupled) until their mass was of the order of the temperature, had the effect of smoothing the anisotropies, i.e. they caused a suppression in the power spectrum of the galaxies that is measured today.
Qualitatively, light neutrinos traveled relativistically for a long period and therefore delayed the formation of structures characterized by a scale smaller than that of the horizon at the time they became non-relativistic. 
The more massive the neutrinos are, the earlier they became non-relativistic, the smaller the scale of the horizon was at that time, inside which the perturbations were smoothed, the more suppressed are the large momenta of the LSS power spectrum.

In formul\ae,  the effect is usually expressed in terms of the quantity $\Omega_\nu$, which is related to the sum of the neutrino masses
\begin{equation}
\Omega_\nu h^2 = \frac{\hbox{Tr}[m\cdot\varrho]}{93.5 \eV}
\label{eq:lss}
\end{equation}
where $m$ is the $4\times4$ neutrino mass matrix and $\varrho$ is the $4\times4$ 
neutrino density matrix (at late cosmological times), discussed and computed above. In a standard case, the numerator corresponds to $\sum m_{\nu_i}$. 
In general, the determination of $\Omega_\nu$ depends on priors and on normalisations, possibly fixed by the CMB spectrum.
As a rule of thumb, the present bound~\cite{boundMnu} and the future expected sensitivity~\cite{future boundMnu} can be summed in 
\begin{equation}
{\rm present:} \qquad \Omega_\nu h^2 \circa{<} 10^{-2}
\end{equation}
\begin{equation}
{\rm future:} \qquad \Omega_\nu h^2 \circa{<} 10^{-3} .
\end{equation}

\medskip

{\bf CMB:} Finally, neutrinos can be studied through the pattern of the CMB anisotropies measured by WMAP (and other experiments). They affect the CMB anisotropies in various ways~\cite{nu in CMB}; the all important quantity is their contribution to the relativistic energy density
\begin{equation}
{\rm Tr} [\varrho] = \rho_{\nu_{\rm tot}} \subset \rho_{\rm rel}
\end{equation}
(where again $\varrho$ is the $4\times4$ neutrino density matrix (at late cosmological times), discussed and computed above), straightforwardly parameterized in terms of an effective number~\footnote{Actually, the naive $N_\nu$ discussed at the beginning of the Section is usually defined as equal to this $N_\nu^{\rm CMB}$. It should be clear from the above discussion that, instead, $N_\nu^{^4{\rm He}}$ and $N_\nu^{\rm D}$ stand for two other different quantities. The three of them come to coincide only in the limiting case in which the only effect of the extra degrees of freedom is a contribution in the total energy density.} of neutrinos $N_\nu^{\rm CMB}$
 \begin{equation}
 \label{eq:CMB}
 \rho_{\rm rel} = \rho_\gamma\bigg[1 + \frac78\bigg(\frac{4}{11}\bigg)^{4/3} N_\nu^{\rm CMB}\bigg].
 \end{equation}
Global fits at the moment imply~\cite{Nnu CMB}
\beq N_\nu^{\rm CMB} \approx 3\pm2\eeq
somewhat depending on which priors and on which data are included in the fit.
Future data might allow a better discrimination.

\subsection{Results}
\label{results cosmology}


\begin{figure}[t]
$$\hspace{-8mm}\includegraphics[width=18cm]{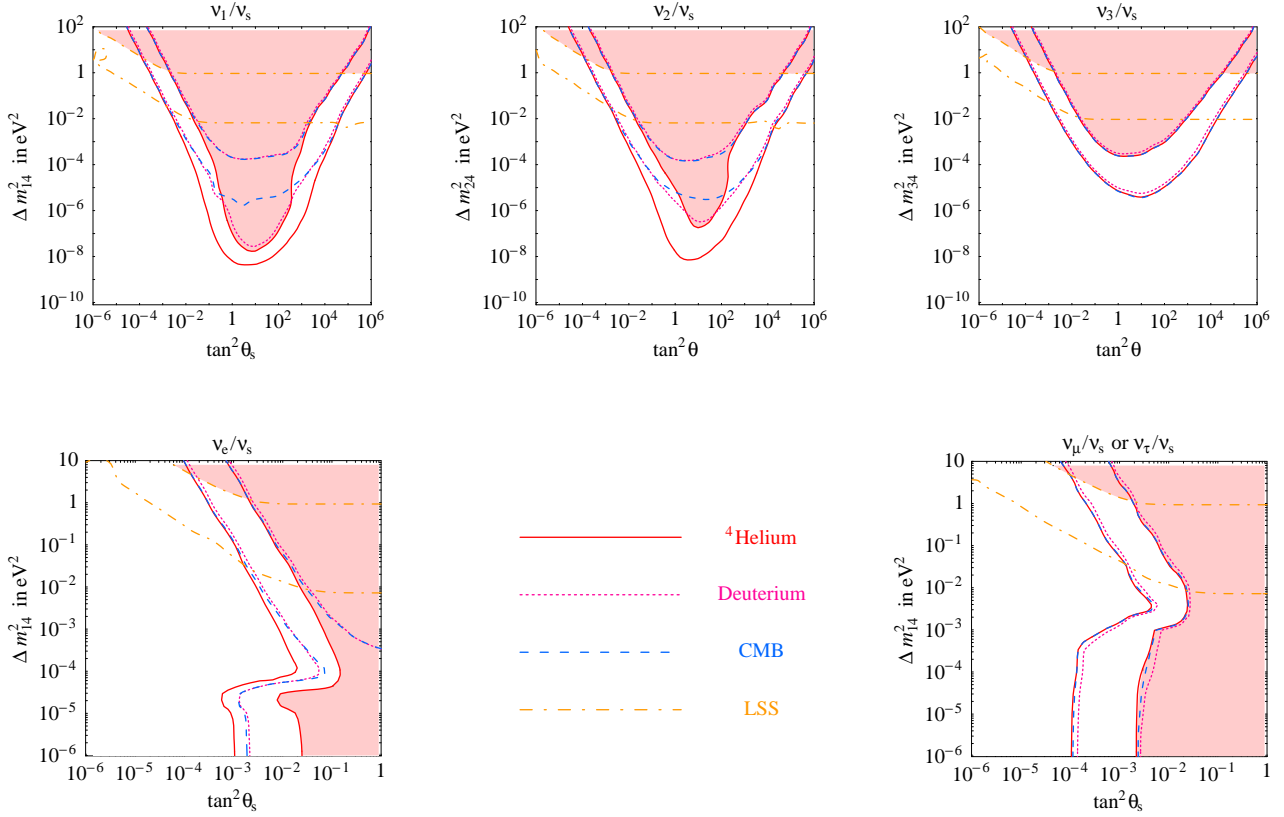}$$
\caption{\label{fig:cosmology}
Cosmological effects of sterile neutrino oscillations. 
\em We collect four different signals.
{\color{rosso} The continuous red line refers to the $^4\hbox{\rm He}$ abundance}
(we shaded as `strongly disfavoured' the regions where its value corresponds to $N_\nu^{^4{\rm He}} > 3.8$ i.e. $Y_p\circa{>}0.26$),
{\color{viola} the purple dotted line to the deuterium abundance}, and
{\color{blu} the dashed blue line to the effective number of neutrinos at recombination}.
We plotted isolines of these three signals corresponding to an effective number of
neutrinos $N_\nu=3.2$ and $3.8$.
The precise meaning of the parameter $N_\nu$ in the three cases is  explained in the text.
{\color{rossoc}
The upper (lower) dot-dashed orange lines corresponds to $\Omega_\nu h^2= 10^{-2}$ $(10^{-3})$}; we shaded as `strongly disfavoured' by the data the regions where $\Omega_\nu h^2 > 10^{-2}$.}
\end{figure}

The results are collected in Fig.\fig{cosmology}.
{\bf We plot} the effective numbers $N_\nu$ of neutrinos which translate the physical observables (the $^4$He and D abundances and the energy density in neutrinos 
at recombination), as dictated by eqs.\sys{HeD} and eq.\eq{CMB}, and the value of the quantity $\Omega_\nu h^2$, defined by eq.\eq{lss}. 
We shade the regions that correspond to $Y_p \circa{>} 0.26$ (i.e. $N^{^4{\rm He}}_\nu> 3.8$)
or $\Omega_\nu h^2>10^{-2}$ and are therefore `strongly disfavoured' or `excluded'
(depending on how conservatively one estimates systematic uncertainties)
within minimal cosmology. The other lines indicate the sensitivity that future experiments might reach.

In order to {\bf qualitatively understand} these precisely computed results, it is useful to begin with the case of mixing with a matter eigenstate $\nu_{1,2,3}$ (upper row of Fig.\fig{cosmology}), and to consider first the line which corresponds to $N_\nu^{\rm CMB}$ (blue dashed line): indeed, this quantity is sensitive only to the total number of neutrinos, and not to the specific flavor ($\nu_e$, $\nu_\mu$ or $\nu_\tau$) which mixes with $\nu_s$. 
In the region above the lines, the production of sterile neutrinos via oscillations is efficient and populates to some extent the sterile species. The slope of the lines can be reproduced by simple analytical estimates~\cite{bbnosc}, and is also intuitive: loosely speaking, at small $\Delta m^2$ the production starts late and needs a large mixing angle to be efficient enough; at large $\Delta m^2$, on the contrary, even a small mixing angle (i.e. a small rate of production) gives rise to a significant amount.
Effects are larger at $\theta_s > \pi/4$ (i.e.\ $\tan\theta_{\rm s}>1$) because this corresponds
to having a mostly sterile state lighter than the mostly active state,
giving rise to MSW resonances (both in the neutrinos and anti-neutrinos channels).

In the region of $\Delta m^2 \circa{<} 10^{-5}\eV^2$, the production starts {\em too} late, namely after neutrino freeze-out. The only effect of the oscillations is then to redistribute the energy density between the active and sterile species, keeping the total entropy constant (there can be no ``refill'' from the thermal bath) so the bound on $N_\nu^{\rm CMB}$ does not apply.

Let us then move to discuss the BBN probes ($^4$He and D abundances): at large $\Delta m^2$ their isolines essentially coincide with $N_\nu^{\rm CMB}$.
On the contrary, in the region of $\Delta m^2 \circa{<} 10^{-5}\eV^2$, if $\nu_e \to \nu_s$ oscillations occur then $\nu_s$ are created by depleting $\nu_e$, as just discussed. Thus the $n/p$ ratio is affected and consequently the $^4$He abundance and, to a lesser extent, the D abundance.
This is apparent in the $\nu_1/\nu_s$ and $\nu_2/\nu_s$ plots of fig.\fig{cosmology}, where the bound extends in the lower part. The bound is stronger for the eigenstate which contains the larger portion of electron flavor: $\nu_1$. In the case of $\nu_3/\nu_s$ mixing nothing happens because no significant $\nu_e$ component is present in $\nu_3$, as $\theta_{13}$ is very small.\\
In turn, for $\Delta m^2 \circa{<} 10^{-8}\eV^2$ oscillations begin {\em really too} late, namely even after the decoupling of $n\leftrightarrow p$ reactions. At that stage, the relative $n/p$ abundance is no more affected by the neutrino populations (but only by neutron decay), so no bounds apply.

The mixing with flavor eigenstates $\nue, \num, \nut$ (lower row of Fig.\fig{cosmology}) is qualitatively different, for the general reasons explained in section~\ref{4nu}.
Essentially, since the flavor eigenstate is spread on two or three mass eigenstates, the active $\to$ sterile oscillations occur at two or three different $\Delta m^2$, of which one is always large enough to give an effect for every $\Delta m_{14}^2$ (our variable of choice for the vertical axis). 

In the case of $\nu_e/\nu_s$ mixing, we see the effect of the solar mass splitting as a bump in the corresponding panel in fig.\fig{cosmology}: if the (mainly) sterile state lies below $\nu_2$, the $\Delta m^2_{\rm sun} \sim 7~10^{-5}\eV^2$ becomes the dominant mass difference.\footnote{Keep in mind that we are assuming a negligible $\theta_{13}$ so there is no electron component in the third mass eigenstates which can feel the mixing with the sterile neutrino. That's why $\Delta m^2_{\rm atm}$ has no role in this case.} The bounds then approximate to vertical lines and they are stronger in $\tan \theta_s$ because of the resonant disposition. More precisely, this is true for the moderate effect expressed by the $N_\nu = 3.2$ line; $\Delta m^2_{\rm sun}$ is sufficient to cause a more incisive effect ($N_\nu = 3.8$ or more) only in the case of $^4$He.\\
In the case of  $\nu_{\mu,\tau}/\nu_s$ mixing, the $\Delta m^2_{\rm atm}$ plays exactly the same role, and this time it is large enough to affect all probes. 


Finally, the LSS structure bound on $\Omega_\nu h^2$ consists of an horizontal line in the central part of the plots: qualitatively, as long as the sterile species is fully populated it is its contribution to the sum of the masses (i.e. the $\Delta m^2$) which is bounded from above. The constraints get weaker for very small $\tan \theta_s$, where the sterile neutrinos are less efficiently produced ($\rho_{\nus} \ll $). 
At $\tan \theta_s > 1$ the bound from $\Omega_\nu$ holds even for very small mixing, $\theta_s \simeq \pi/2$, just because this region corresponds to heavy active neutrinos.

{\bf In summary}, fig.\fig{cosmology} displays the excluded, the allowed and the future testable regions of the active/sterile mixing parameter space, for six limiting cases, for what concerns cosmological probes. Any specific model of a sterile neutrino identifies a preferred point (or area) in one of these spaces or in a suitable combination of them, and should therefore be tested on them.
This is what we do, as an example, for the LSND sterile neutrino in the next Section.

\subsection{LSND: in or out?}
\label{LSND section}

\begin{figure}
$$\includegraphics[width=7cm]{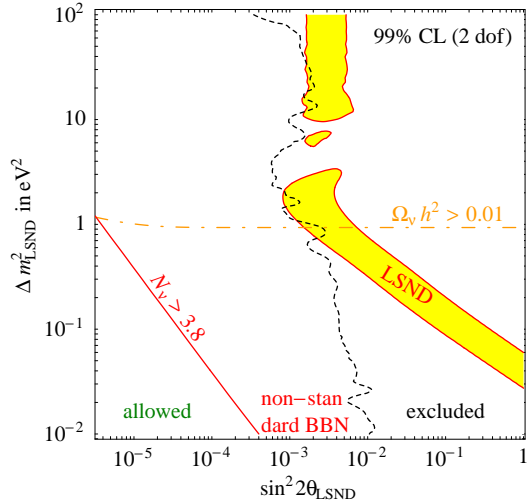}$$
\caption{\label{fig:LSND} 
The LSND anomaly interpreted as oscillations of 3+1 neutrinos versus the cosmological constraints.
\em Shaded region: suggested at 99\% C.L.\ by LSND.
Black dotted line: 99\% C.L.\ global constraint from other neutrino experiments
(mainly Karmen, Bugey, SK, CDHS).
{\color{rossos} Continuos red line: the bound from BBN corresponding to $Y_p \simeq 0.26$, i.e. $N_\nu = 3.8$ thermalized neutrinos.}
{\color{rossoc} Dot-dashed orange line: the bound from LSS corresponding to $\Omega_\nu h^2 =0.01$.}}
\end{figure}

The LSND experiment reports a signal \cite{LSND} for $\numb \to \nueb$ oscillations in the appearance of $\nueb$ in an originally $\numb$ beam. The best fit point is located at $\sin^2\theta_{\rm LSND} = 3 \cdot \, \, 10^{-3}$, $\Delta m^2_{\rm LSND} = 1.2 \eV^2$ (the whole allowed region is represented in fig.\fig{LSND}). 

The ``3+1 sterile'' neutrino explanation assumes that the $\numb \to \nueb$ oscillation proceeds through $\numb \to \nusb \to \nueb$. The large LSND mass scale $\Delta m^2_{\rm LSND}$ separates the three active states (split by the solar and atmospheric gaps) from the additional sterile state. The effective angle of the LSND oscillation $\theta_{\rm LSND}$ can be expressed in terms of the two active-sterile angles $\theta_{es}$, $\theta_{\mu s}$ as $\theta_{\rm LSND} \approx \theta_{es} \cdot \theta_{\mu s}$.\footnote{In our parameterization, this simply corresponds to a unit vector $\vec n = (n_e, n_\mu, n_\tau) \simeq (\frac{1}{\sqrt{2}},\frac{1}{\sqrt{2}},0)$.}
However, each one of these two angles is constrained by several other experiments that found no evidence of electron or muon neutrino disappearance. Moreover, the $\numb \to \nueb$ oscillations are directly excluded by KARMEN in a fraction of the parameter space. As a result \cite{strumia giunti lsnd}, a large portion, but not the totality, of the area indicated by the LSND experiment is ruled out (see fig.\fig{LSND}). The poor compatibility with solar and atmospheric oscillation data, in the context of four $\nu$ mixing, also puts the LSND sterile neutrino in a difficult position~\cite{maltoni 4nu}.

How does the LSND signal compare to the cosmological bounds discussed above? Fig.\fig{LSND} shows the constraints from BBN and from LSS superimposed to the LSND region\footnote{In constructing fig.\fig{LSND} the BBN constraint has been minimized setting $\theta_{es} \approx \theta_{\mu s} \approx \theta_{\rm LSND}$, when this is allowed by the neutrino disappearance data.}. We see that {\bf the entire LSND region is ruled out by the BBN constraint}: basically, for every value of its mixing parameters the LSND sterile neutrino completely thermalizes and implies an unacceptable modification of the $^4$He primordial abundance.
Of course, remember that allowing non-standard modifications to cosmology/BBN has the power to relax the bound to some extent.  Exemplar is the case of a large primordial neutrino asymmetry, as discussed above. Other recent suggestions include~\cite{non standard nu in cosmology}.\\
The constraint on $\Omega_\nu$ is well approximated by the horizontal line
corresponding to $\Omega_\nu h^2 = m_4/93.5\eV$. It starts to bend (as discussed in Sec.\ref{results cosmology}) only at smaller values of the effective $\theta_{\rm LSND}$ mixing angle.

Fig.\fig{LSND}, that combines the precise computation of the evolution of mixed neutrinos in the Early Universe and of the neutrino experimental data, therefore reproduces and completes the estimates already presented in~\cite{LSNDBBN} and~\cite{LSNDOmegaNu}.

\section{Sterile Neutrinos in Supernova sauce}
\label{SN}

{\bf Generalities:} Supernov\ae\ can be powerful and important laboratories for neutrino physics, and therefore in particular for the physics of sterile neutrinos. The fundamental reasons for this basic fact are simply listed:
\begin{itemize}
\item SNe are abundant sources of neutrinos, since this is the main channel into which most of their enormous energy is emitted; as a consequence, neutrinos play a crucial role in the evolution of the SN phenomenon; 
\item given the characteristic temperatures of the SN environment, the typical energy of the emitted neutrinos ($\sim10\div20\MeV$) is such that they can be easily detected on Earth; 
\item SNe are so far away that neutrinos must travel over distances so large that they have plenty of time (or space) to experience and fully develop the consequences of several ``exotic'' effects (oscillations, non-conventional very feable interactions, decay...), if any is present;
\item SN cores are so extremely dense that neutrinos remain trapped and undergo matter effects that cannot be relevant anywhere else.
\end{itemize} 

On the other hand, it is true that the physics of supernov\ae\ is very complicated and demanding, and could pose a threat on their possible usefulness as ``clean experiments''. Nevertheless, the basic features are robust enough to be used as incontrovertible criteria, sometimes, maybe, requiring a sensible compromise between detailness and usefulness in the treatment of SN physics. 
At the present epoch, the twenty-something events of the SN1987a signal~\cite{SN1987a} already allow to  set cautious constraints. 
The future is however brighter: running solar neutrino experiments could detect  thousands of events 
from a future SN  exploding at distance $D\sim 10$~kpc and an even more impressive harvest of data could come from a future Mton water-\v{C}erenkov detector or from other more SN-oriented future projects~\cite{Cei}, making the quantitative analysis of the SN neutrinos worthwhile for the search of sterile states. Previous analysis have investigated the subject~\cite{sterile SN}.

{\bf Neutrino evolution:} What we need to do is simply said: we must follow the fate of the neutrinos emitted from neutrino-spheres\footnote{Neutrino-sphere: the region of the star mantle at which the density becomes low enough that neutrinos (produced in the core) are no more trapped and freely stream outwards.} along their travel through the star mantle, the vacuum and (possibly) the Earth.~\footnote{For the range of $\Delta m_s^2\ \circa{<}\ {\rm few}\ \eV$ to which we are confined by the cosmological bounds discussed in Sec.\ref{cosmology}, the MSW resonances with the sterile state occur out of the neutrino-sphere. The resonances would enter in the neutrino-spheres (in the inner core) for $\Delta m_s^2 \circa{>} 10^5 \eV^2$ ($\circa{>} 10^7 \eV^2$ respectively).}
The existence of the sterile neutrino can introduce modifications at each of these steps, via matter or vacuum conversions, in different ways for each possible choice of the sterile mixing parameters.

{\footnotesize
In more detail: one has to follow the evolution of the $4\times 4$ neutrino density matrix $\varrho_m$, written in the basis of instantaneous mass eigenstates. For instance, a $\nu_e$ with energy $E_\nu$ is described by
 $\varrho_m  = V_m^\dagger\cdot
\hbox{diag}\,(1,0,0,0)\cdot V_m $
where $V_m$ depends on $E_\nu$ and, in general, on the position in the star.
The mixing matrices in matter ($V_m$) and vacuum ($V$)
are computed diagonalizing the Hamiltonian
\begin{equation}
\label{eq:Ham SN}
\mathcal{H} = \frac{1}{2E_\nu}\left[ V \, {\rm diag}(m_1^2,m_2^2,m_3^2,m_4^2) \, V^\dagger  + E_\nu \textrm{diag}(\Ve,\Vm,\Vt,0) \right]
\end{equation}
and ordering the eigenstates according to their eigenvalues $H_i \equiv m_{\nu_{mi}}^2/2E_\nu$:
$\nu_{m1}$ ($\nu_{m4}$) is the lightest (heaviest) neutrino mass eigenstate in matter. 
The evolution up to the detection point is described by
a $4\times4$ unitary evolution matrix $\mathcal{U}$ so that
at detection point the density matrix $\varrho$ in the basis of flavor eigenstates is
\begin{equation}
\label{eq:matrix evolution}
\varrho = V\cdot \mathcal{U}\cdot\varrho_m(E_\nu)\cdot \mathcal{U}^\dagger \cdot V^\dagger \qquad {\rm with} \qquad
\mathcal{U} = \mathcal{U} _{\rm Earth}\cdot \mathcal{U} _{\rm vacuum}\cdot \mathcal{U} _{\rm star}.
\end{equation}
The evolution in vacuum is simply given by 
$\mathcal{U} _{\rm vacuum} = \diag\exp (-i L m_{\nu_i}^2/2E_\nu)$.
Combined with average over neutrino energy it suppresses
the off-diagonal element $\varrho_m^{ij}$ when
the phase differences among eigenstates $i$ and $j$ are large.\\
The evolution in the matter of the star is more complicated because several level crossings can occur, say at radii $r_1 \ldots r_N$. At each one of them, there is a certain ``jump'' probability. In the present formalism, this can be expressed by a $4 \times 4$ rotation matrix $P$, with the rotation angle given by $\tan^2\alpha = P_C/(1-P_C)$ where $P_C$ is the level crossing probability. Indeed, in particular, when levels $i$ and $j$ cross in an adiabatic way, $P=\mathbb{I}$. If instead the level crossing is fully non adiabatic $P$ is a rotation with angle $\alpha = 90^\circ$ in the $(ij)$ plane.

The computation of $P_C$ at the crossings requires attention. Focussing e.g. on $\theta_s < \pi/4$, at a crossing between a mainly active state $\nu_a$ and the mainly sterile state $\nu_s$ we compute it as 
\begin{equation}\label{eq:PC}
 P_C = \frac{e^{\tilde\gamma \cos^2\theta^m_{as}}-1}{e^{\tilde\gamma} -1}\qquad
\gamma =\frac{ 4 \mathcal{H}_{as}^2}{d H_a/dr}
\equiv
\tilde\gamma \cdot \frac{\sin^2 2\theta^m_{as}}{2\pi| \cos2\theta_{as}^m|}
\quad\hbox{where}\quad \sin\theta_{as}^m = \vec{n}\cdot \vec{\nu}_a^m \sin\theta_{\rm s} .
\end{equation}
where it is important to notice that $\gamma$ and $\theta^m_{as}$ must be computed
around the resonance, where $\mathcal{H}_{aa}=\mathcal{H}_{ss}$ (or around the point where adiabaticity is maximally violated, in cases where there is no resonance)
and are in general different from their vacuum values (that are instead conveniently used to parameterize $P_C$ in the simpler $2\nu$ case).~\cite{friedland}

Between a level crossing and the following one the evolution proceeds as governed by the matter Hamiltonian, so that the complete form for $\mathcal{U}_{\rm star}$ is
\begin{equation}
\mathcal{U}_{\rm star} = P_{r_n}\cdots P_{r_2}\cdot 
\diag\exp \left( -i\int_{r_1}^{r_2} ds~\frac{ m_{\nu_{mi}}^2}{2E_\nu} \right) \cdot  P_{r_1}\cdot
\diag\exp \left( -i\int_{r_0}^{r_1} ds~\frac{ m_{\nu_{mi}}^2}{2E_\nu} \right).
\end{equation}
In practice, given the importance of the matter effects in the star mantle and the very long baseline to Earth, the evolution between the level crossings and in vacuum averages to zero the off-diagonal elements that are possibly produced by the rotation matrices, introducing significant simplifications. However, if two states have $\Delta m^2\circa{<} 10^{-18}\eV^2$
vacuum oscillations do not give large phases: evolution in the outer region of the SN and in vacuum
must be described keeping the off-diagonal components of the density matrix.\\
Finally, for simplicity we now assume that the neutrinos do not travel through the Earth matter, so that $\mathcal{U} _{\rm Earth}$ is trivial. In the case of the SN1987a signal, this was not true but disregarding it only implies a few percent error, well within the general uncertainties for our purposes. In the case of the next SN event, it could be in general reintroduced.

Having described the general formalism, let us now focus more closely on the peculiarities of the SN case.
At the neutrinosphere, matter effects are dominant, so that
matter eigenstates coincide with flavor eigenstates (up to a trivial permutation dictated by the MSW  potentials that will be more clear below): the initial density matrix consists of $\varrho_m = \diag ( \Fso,\Feo, \Ftauo,\Fmuo)/\Phi_{tot}^0$, where $\Phi^0_{\bar\nu}$ are the flavor fluxes from the neutrinosphere and $\Phi_{tot}$ stands for their sum.
The final fluxes at the detection point will be then given by the diagonal entries of $\varrho$ of eq.\eq{matrix evolution}: $(\Fe,\Fmu, \Ftau,\Fs) = \diag (\varrho) \cdot \Phi_{tot}$.\\
Concerning the initial fluxes, the accurate results of simulations are usually empirically approximated by a Fermi-Dirac spectrum for each flavor $\nue,\nueb, \nu_x$
($\nu_x$ collectively denotes $\nu_{\mu,\tau},\bar\nu_{\mu,\tau}$), with a ``pinching'' that slightly suppresses the low energy portion and the high energy tail.
Based on the recent results of~\cite{keil}, we adopt the following average energies and 
total luminosities for the various neutrino components
at the time of the snapshot of fig.\fig{profiles} (see below): $\langle E_{\nue,\nueb,\nu_x}\rangle\simeq12, 14, 14 \MeV, L_{\nue,\nueb,\nu_x}\simeq30, 30, 20 \cdot 10^{51} \erg \sec^{-1}$, assuming (in accordance with numerical calculations) that the ratios of luminosities do not vary much during the 
whole emission.
The initial flux of sterile neutrinos is assumed to be vanishing, as a consequence of the fact that matter oscillations only take place out of the neutrinosphere.
\begin{figure}[t]
\begin{center}
\includegraphics[width=16cm]{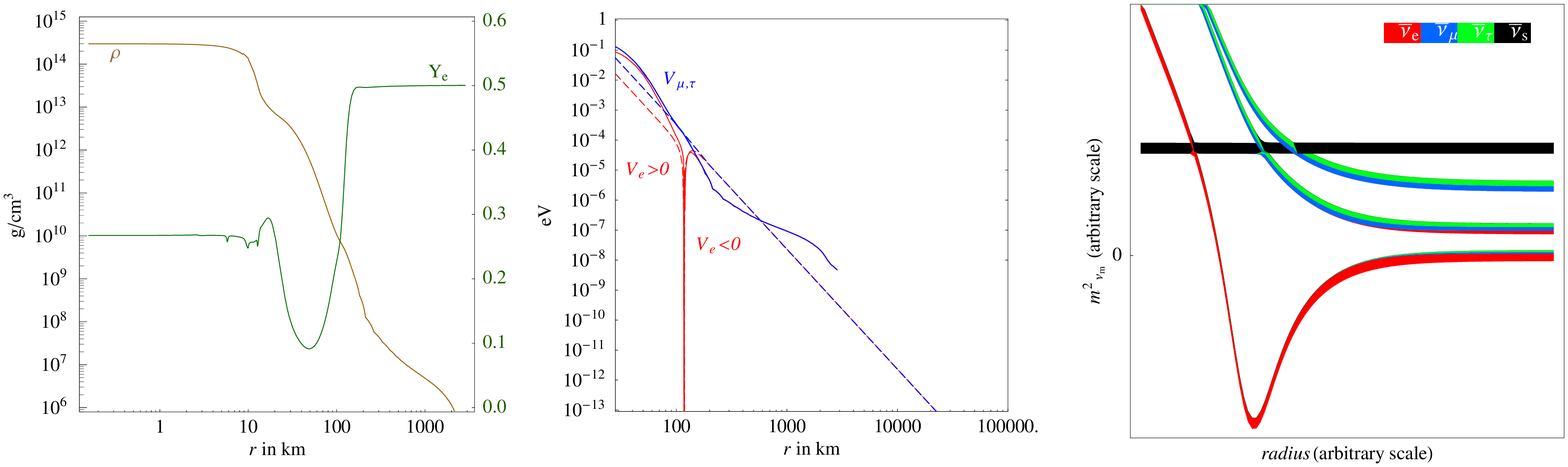}
\caption{\label{fig:profiles} Inside the supernova.
\em Left panel:
{\color{rosso} Density $\rho(r)$} and {\color{verdes} electron fraction $Y_e(r)$} from~\cite{burrows}.
Center panel: 
matter potentials for {\color{rossos} $\bar\nu_e$} (red solid line) and for {\color{blu} $\bar{\nu}_{\mu,\tau}$} (blue solid line); the dashed lines are the analytic modelization that we adopt.
Right panel: 
a representation of the antineutrino matter eigenstates with their flavor content, as functions of the radius in the interior of the SN; the drawing is made in the specific case of small $\nu_e/\nu_s$ mixing and large  $\Delta m^2_{14} (> \Delta m^2_{\rm atm})$ for the sake of illustration; we are assuming small $\theta_{13}$ and normal hierarchy.
}
\end{center}
\end{figure}

The MSW potentials of eq.\eq{Ham SN}, experienced by the neutrinos in SN matter are
\beq\label{sys:potentials}\begin{array}{ll}
V_e =\sqrt{2} \GF \nB \left( 3 Y_e - 1\right)/2,\qquad  & V_\tau =V_\mu + V_{\mu\tau}, \\
\displaystyle
V_\mu = \sqrt{2} \GF \nB \left({Y_e} -1 \right)/2,  & V_s = 0, \end{array}
\eeq
where $\nB$ is the baryon number density ($\nB =
{\rho(r)}/{m_N}$ where $m_N \approx 939 \MeV$ is the nucleon mass
) 
and $Y_e=(N_{e^-} - N_{e^+})/\nB$ 
is the  electron fraction per baryon.
Antineutrinos experience the same potentials with opposite sign.
The difference $V_{\mu\tau}$ in the $\num$ and $\nut$ potentials, which appears at one
loop level due to the different masses of the muon and tau leptons \cite{mutau}, is,
according to the SM
\begin{equation}
V_{\mu \tau}=\frac{3\GF^2 m^2_{\tau} }{2 \pi^2} \left[ 2 (n_p + n_n) \ln \left(
\frac{M_W}{m_\tau} \right) -n_p -\frac{2}{3}n_n \right].
\end{equation}
The effect is not irrelevant in the inner dense regions: for densities above $\rho \sim 10^{8} \gcm$, the $\mu\tau$ vacuum mixing is suppressed.

A crucial point concerns the characteristic of the matter density $\rho(r)$ and of the electron fraction $Y_e$ in the mantle of the star. We adopt the profiles represented in fig.\fig{profiles} and we model them with analytic functions that preserve their main features.
Namely, the density profile decreases according to a power law $r^{-4}$ out of the $\sim 10 \km$ inner core (where instead it has a roughly constant, nuclear density value). 
At much larger distances the density profile gets modified in a time-dependent way by the
passage of the shock wave.
Present simulations have difficulties in reproducing this phenomenon
and therefore cannot reliably predict the density profile in the outer region.
Therefore for $r\circa{>}500\km$ 
we assume a power law $\rho(r) = 1.5~10^4(R_\odot/r)^{3}{\rm g/cm}^3$, which roughly
describes the static progenitor star.\\
In turn, the peculiar $Y_e$ profile in fig.\fig{profiles}
is inevitably dictated by the deleptonization process: behind the shock wave which has passed in the mantle matter, the electron capture on the newly liberated protons is rapid, driving $Y_e$ to low values ($\sim 1/4$). 
In the outer region, where the density is sensibly lower, the efficiency of the capture is much lower, so that $Y_e$ essentially maintains the value $\sim 1/2$ typical of normal matter.~\footnote{The data refer to $\sim$0.3 sec after bounce for a typical star of $\sim 11$ solar masses. 
The subsequent evolution is supposed to move the wave of the $Y_e$ profile slightly outwards, maintaining, however, its characteristic shape.}
This is important because the matter potential $V_e$ of electron (anti)neutrinos 
flips sign, see eq.~(\ref{sys:potentials}), when, in the deep region of the mantle $Y_e$ steeply decreases below $1/3$.

  


The knowledge of the matter potentials (and of the neutrino masses and mixings) allows to draw the pattern of the (anti-)neutrino eigenstates with their flavor content, as functions of the radius $r$ in the interior of the star. 
This is represented in the right panel of fig.\fig{profiles}, in the specific case of small $\nu_e/\nu_s$ mixing and large  $\Delta m^2_s (\gg \Delta m^2_{\rm atm})$ for the sake of illustration.
}

In the end, we collect the modified (anti)neutrino fluxes and deduce the modifications to some relevant observables; in particular, we focus on the final flux of $\nueb$, which are best detected through $\nueb p \to e^+ n$ at the \v{C}erenkov detectors. 

\subsection{Results}

\begin{figure}[!ht]
$$\includegraphics[width=15.8cm]{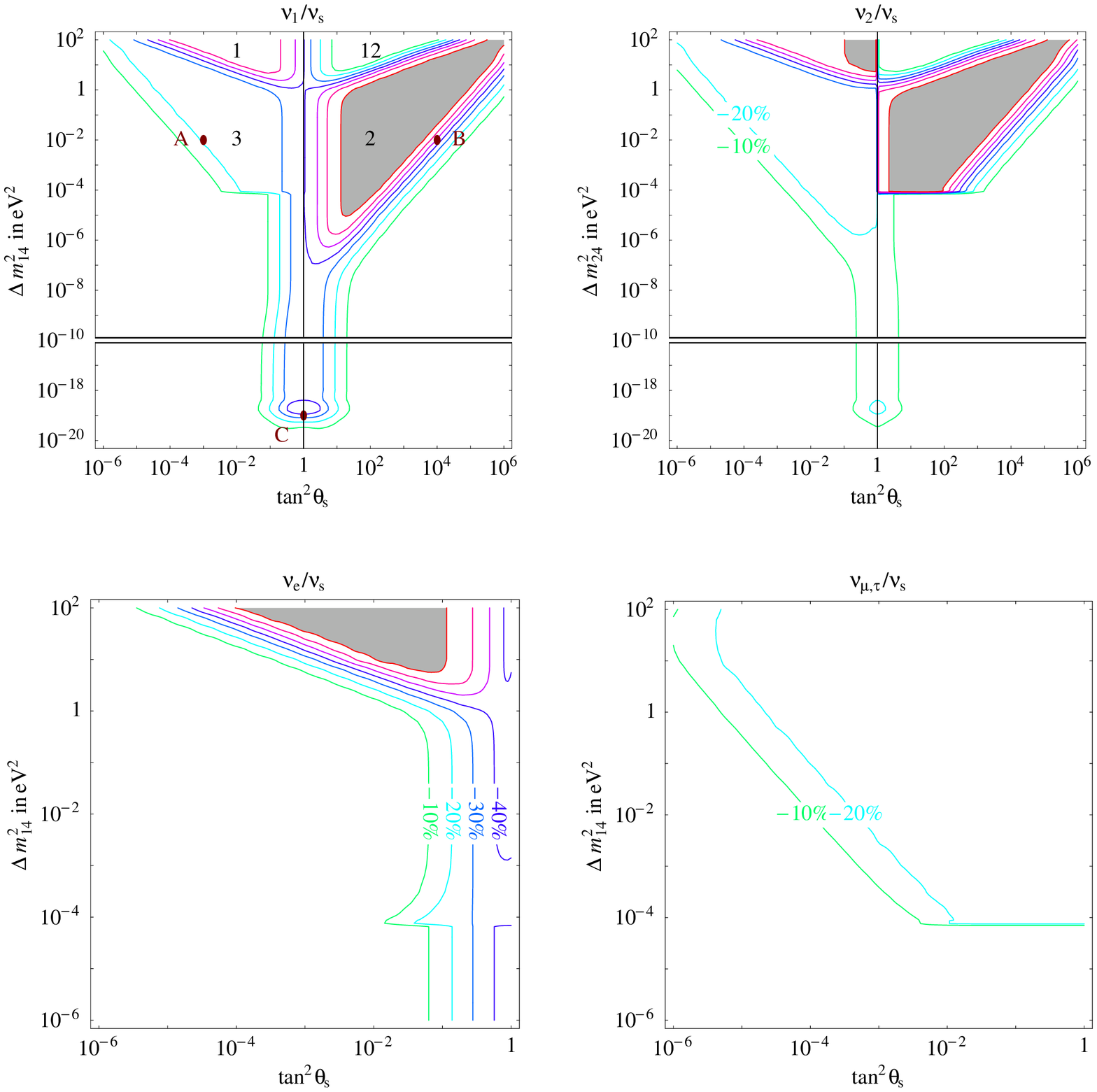}$$
\caption{\label{fig:SN} Sterile effects in supernov\ae.
\em The iso-contours correspond to a ${\color{verdes}10}$, ${\color{blucc}20},
{\color{blu}30}$, ${\color{blus}40},{\color{viola}50}$, ${\color{viola2}60},{\color{rossos}70}\,\%$ deficit
of the SN $\bar\nu_e$ total rate due to oscillations into sterile neutrinos.
The deficit is measured with respect to the rate in absence of
active/sterile oscillations but of course in presence of active/active oscillations
(which reduce the unrealistic no-oscillation-at-all rate by $\sim 10\%$).
We shaded as disfavoured by SN1987a data regions with
a deficit larger than $70\%$.
While the qualitative pattern is robust,
regions with MSW resonances can shift by one order of
magnitude in $\Delta m^2$ using different SN density profiles.
$\bar\nu_3/\bar\nu_{\rm s}$ mixing (not plotted) does not give significant effects. 
Fig.\fig{SNsample} studies in detail the sample points here marked as {\rm A, B, C},
and the regions 1, 2, 12, 3 are discussed in the text.}
\end{figure}

\begin{figure}[h]
$$\includegraphics[width=0.6\textwidth]{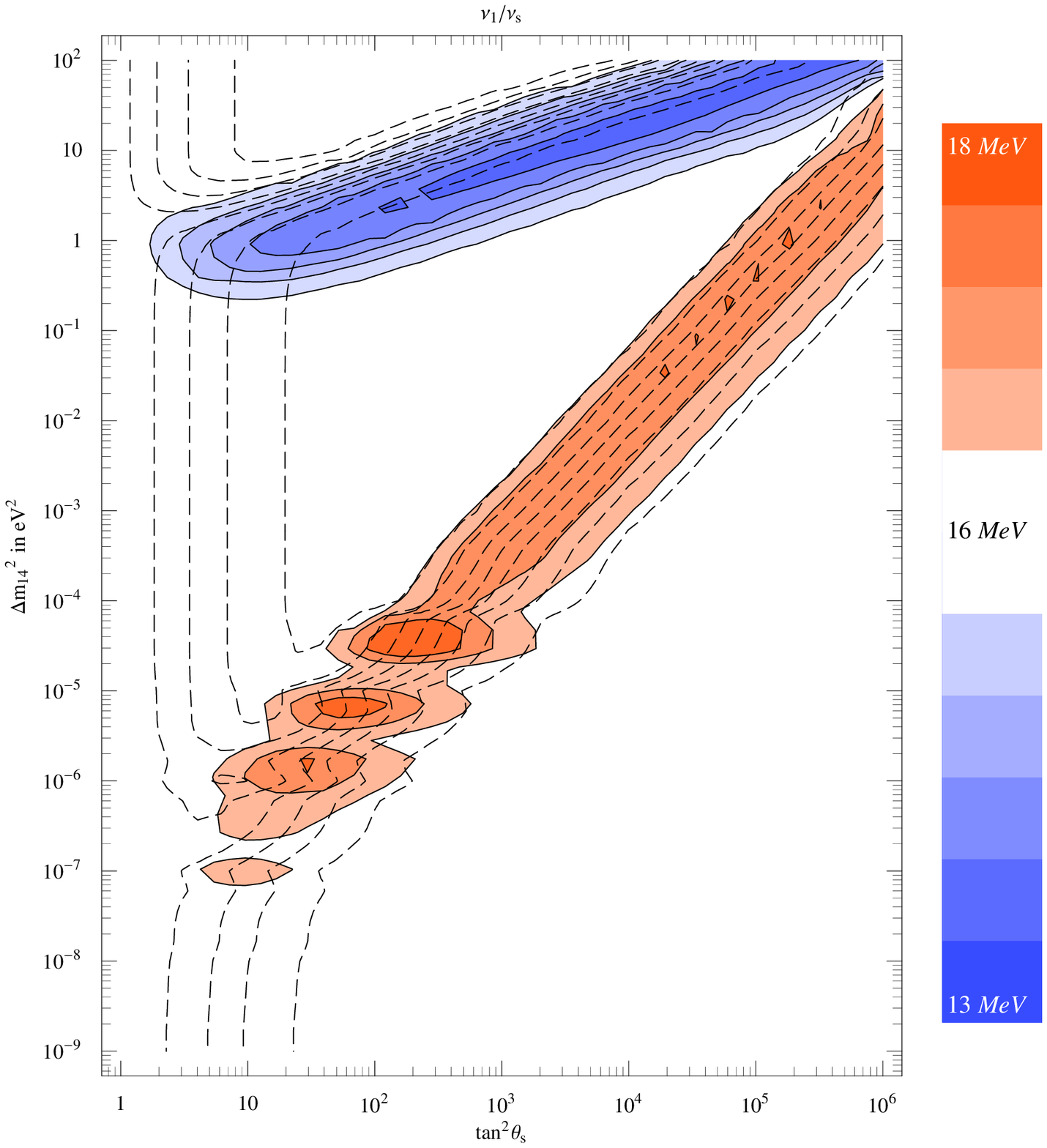}$$ 
$$\includegraphics[width=\textwidth]{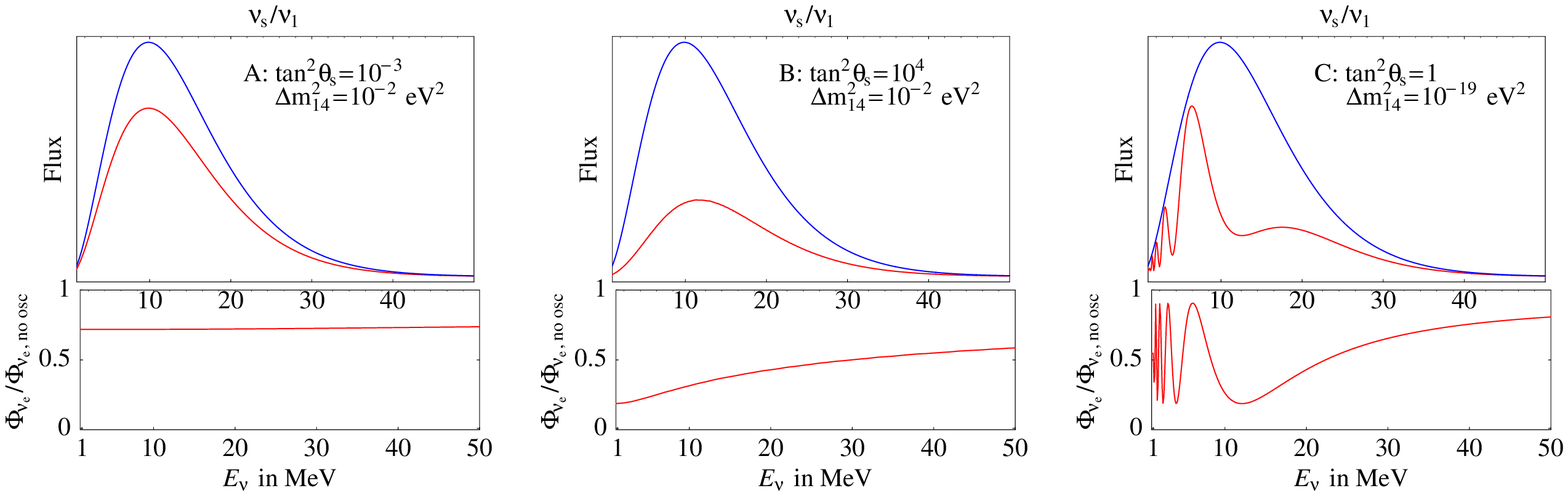}$$
\vspace{-1cm}
\caption{\label{fig:SNsample} Sterile effects in supernov\ae.
\em Average $\nueb$ energy in the $\nu_1/\nu_s$ plane ($\tan^2\theta_{\rm s} > 1$ portion) and distortion of the $\bar\nu_e$ flux at sample points {\rm A, B, C}.}
\end{figure}

The results are collected in fig.\fig{SN}, where we plot the reduction of the $\bar\nu_e$ event rate in a typical \v{C}erenkov detector due to sterile mixing~\footnote{We focus on $\bar\nu_e p$ scatterings with the cuts and efficiency of the KamiokandeII experiment. The cross section is taken from~\cite{IBD}.}, and in fig.\fig{SNsample}, where we plot the modified average energy and the distortion of the $\bar\nu_e$ spectrum for a few specific choices of mixing parameters.

In order to {\bf understand qualitatively} the main features of these results it is useful to look at the pattern of level crossings, like the one qualitatively depicted in fig.\fig{profiles}. Indeed, as discussed above, the {\bf active/sterile MSW resonances} in the matter of the star are the crucial places where the neutrino flux is non-trivially modified, the rest is simply vacuum oscillations. 
There are three possible resonances:
\begin{itemize}
\item[1.] The mostly $\bar\nu_s$ eigenstate crosses the mostly $\bar\nu_e$ eigenstate 
at $r\sim 100\km$, where $V_e$ flips sign.
At this point matter effects dominate over active neutrino masses,
so that active mass eigenstates coincide with flavor eigenstates.
Since $V_e$ flips sign in a steep way this resonance is effective
only if $\Delta m^2_{14}\circa{>} 10^{-1\div0}\eV^2$
(different SN simulations gives values in this range).

\item[2.] If the mostly sterile eigenstate is the lightest one
(in our parameterization this needs $\theta_{\rm s}\circa{>} \pi/4$)
the two eigenstates in 1.\ cross again at larger $r$.
Pictorially,  this second resonance is present when 
the sterile black line is lowered below the others in fig.\fig{profiles}.
This MSW resonance occurs at large $r$  where $V_e$ is smooth,
so that it is effective down to $\Delta m^2_{14}\circa{>} 10^{-6\div 8}\eV^2$.
Again, the significant uncertainty is due to uncertainties on the SN density gradient.

\item[3.] When the mostly sterile eigenstate is
the heaviest or the next-to-heaviest state, it also crosses
one or both of two mostly $\bar\nu_{\mu,\tau}$ eigenstates.
This is the case illustrated in fig.\fig{profiles}.
The values of $\Delta m^2_{24}$ and $\Delta m^2_{34}$
 determine at which $r$ these crossings takes place,
 and consequently the  flavor composition of the mostly active states at the resonance.
 Entering in the SN, the small $\bar\nu_e$ component of $\bar\nu^m_{2,3}$
 disappears as soon as 
$V_e-V_\mu$ dominates over $\Delta m^2_{\rm sun}$.
In any case, active/sterile MSW resonances with the mostly $\bar\nu_{\mu,\tau}$ states affect only marginally the $\bar\nu_e$ rate, right because the $\nu^m_{2,3}$ contribution to $\nueb$ is secondary.
\end{itemize}
At each of the resonances, part of the neutrino flux can then convert into the sterile state and deplete the final active flavors. Combinations of more than one resonance can occur, depending on the mixing parameters. These considerations allow to understand the regions of $\nueb$ {\bf flux reduction} in fig.\fig{SN}.

Let us start from the case of $\nu_{1}/\nu_s$ mixing.
Resonance 1 gives a sizable reduction in region 1 
and resonance 2 gives a sizable reduction in region 2.
Had we ignored solar mixing the maximal deficit would have been $100\%$,
while in presence of solar oscillations the
maximal effect is a $\sim 80\%$ deficit.
More precisely, in the interior of region 1 one obtains
$\Phi_{\bar\nu_e} = \sin^2\theta_{\rm sun} \Phi^0_{\bar\nu_e}$
because resonances 1 and 3 are fully adiabatic.
In the interior of region 2 one obtains
$\Phi_{\bar\nu_e} = \sin^2\theta_{\rm sun} \Phi^0_{\bar\nu_{\mu,\tau}}$
because resonance 2 is fully adiabatic and resonance 1 irrelevant.
Therefore, given the assumed initial fluxes, 
the $\bar\nu_e$ rate gets reduced  slightly more strongly in region 2
than in region 1.\\
In region 12 both resonances 1 and 2 are effective, 
and tend to compensate among each other:
resonance 1 converts $\bar\nu_e$ into $\bar\nu_s$
and resonance 2 reconverts $\bar\nu_s$ into $\bar\nu_e$.\\
In region 3, resonances 3 gives a $20\%$ suppression of the $\bar\nu_e$ rate,
that sharply terminates when $\Delta m^2_{14}< \Delta m^2_{\rm sun}$.
This is due to a strong suppression of the mostly $\bar\nu_{\mu,\tau}$ eigenstates,
which due to solar oscillations would give a $20\%$ contribution to the $\bar\nu_e$ rate
(ignoring solar mixing, there would be no suppression of the $\bar\nu_e$ rate in region 3).
This reduction of the $\bar\nu_{\mu,\tau}$ fluxes induced by resonances 3 
could be better probed by measuring the NC rate (which gets a $\circa{<}40\%$ reduction)
and, if neutrinos cross the Earth, by distortions of the $\bar\nu_e$ energy spectrum.\\
The tail at smaller $\Delta m^2$ and around maximal mixing is due to
vacuum oscillations, that can reduce the $\bar\nu_e$ rate by $\circa{<}50\%$:
their effect persists down to  $\Delta m^2\sim E_\nu/D\sim 10^{-18}\eV^2$.
The precise value depends on the distance $D$
(we assumed $D=10$ kpc).

The other mixing cases are understood in similar ways.
We remind that our parametrization is discontinuous at $\theta_s=\pi/4$:
this is reflected in the panel of fig.\fig{SN} which illustrates $\nu_{2}/\nu_s$ mixing.
Resonance 2 sharply terminates when $\Delta m^2_{24}<\Delta m^2_{\rm sun}$.
Except for these differences, this case is quite similar to the previous one, 
because $\bar\nu_1$ and $\bar\nu_2$ get strongly mixed by matter effects.
On the contrary $\nu_{3}/\nu_s$ mixing (not shown) does not give a significant reduction
of the $\bar\nu_e$ rate.

Mixing with the flavor eigenstates behaves in a similar way. 
Namely, 
the $\nu_{\mu,\tau}/\nu_{\rm s}$ figure shows the reduction in region 3 due to resonances 3,
and the $\nu_e/\nu_{\rm s}$ figure  shows the reduction in region 1 due to resonance 1.
The additional feature at $\Delta m^2_{14} \sim \Delta m^2_{\rm{sun}}$ is
due to adiabatic conversion (for large sterile angles) between the mostly-sterile state and $\nu_2$, 
that are almost degenerate in this condition. 
For larger $\Delta m^2_{14}$, there is no $\bar\nu_e$ component in $\bar\nu^m_2$ so that the crossing is totally non adiabatic, while for smaller $\Delta m^2_{14}$ the two states are separated. 
Sterile effects persists at all values of $\Delta m^2_{14}$, even if it is small.

%

The reductions highlighted in fig.\fig{SN} have to be compared to the measured flux to set constraints and identify anomalies. This will certainly be fruitful in the future. At present, however, the theoretical uncertainties on the SN evolution and fluxes and the smallness of the SN1987a data set only allow to put conservative constraints: we simply shaded as `disfavoured' regions where sterile effects reduce the $\bar\nu_e$ rate by more than $70\%$.

In the same perspective, important observables in the future SN event will be the {\bf $\nueb$ spectrum}, where distortions could be induced by sterile oscillations, or more generally the {\bf average $\nueb$ energy}.
In absence of sterile oscillations we expect $\langle E_{\bar\nu_e}\rangle \approx 15\MeV$ with a quasi-thermal spectrum.
In presence of sterile oscillations, these observables can be affected in selected regions of the parameter space, as illustrated by figure\fig{SNsample}. The average energy can increase up to $\langle E_{\bar\nu_e}\rangle \approx 18\MeV$ and decrease down to $\approx 12\MeV$ along the sides of the MSW triangle. The distorted spectrum which correspond to the first case is plotted in fig.\fig{SNsample}B.
Vacuum oscillations can also give well known distortions,
as exemplified in fig.\fig{SNsample}C.
These effects seem larger than experimental and theoretical uncertainties.
In all other cases sterile effects give a quasi-energy-independent suppression of
the $\bar\nu_e$ rate, and therefore negligibly affect $\langle E_{\bar\nu_e}\rangle$.
Fig.\fig{SNsample}A gives an example of this situation.


\begin{figure}[t]
$$\hspace{-7mm}\includegraphics[width=17cm]{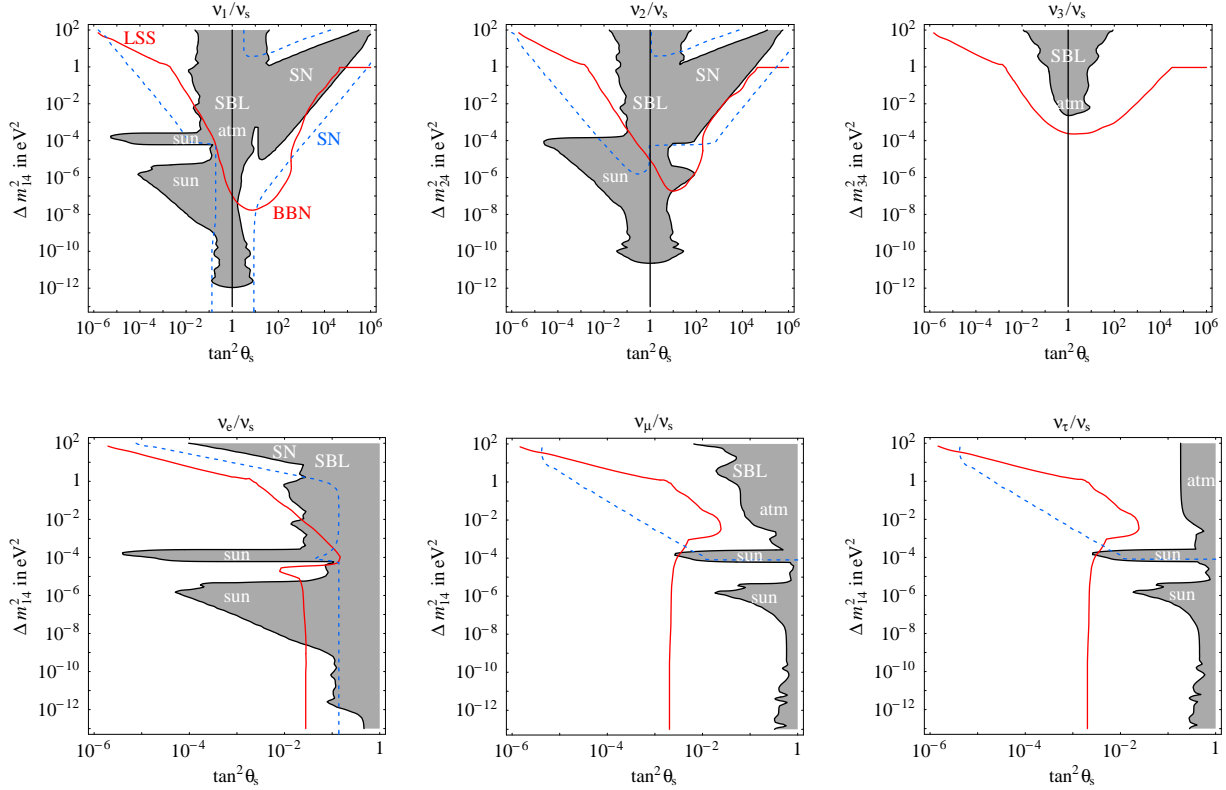}$$
\caption{\label{fig:all} Summary of sterile neutrino effects.
\em The shaded region is excluded at $99\%$ C.L.\ (2 dof)
by solar or atmospheric or reactor or short base-line experiments.
We shaded as excluded also regions where sterile neutrinos
suppress  the SN1987A $\bar\nu_e$ rate by more than $70\%$.
This rate is suppressed by more than $20\%$ inside the {\color{blue} dashed blue line},
that can be explored at the next SN explosion if it will be possible to
understand the  collapse well enough.
Within standard cosmology,
the region above the {\color{rossos} red continuous line} is disfavoured (maybe already excluded) by BBN and LSS.}
\end{figure}

\section{Conclusions}
\label{conclusions}

We have performed in~\cite{CMSV} a systematical study of the effect of an extra sterile neutrino in all possible contexts, for any choice of its mixing parameters with the active neutrinos and fully including the active/active mixings now established by the results on solar and atmospheric oscillations. We considered cosmology (BBN, CMB, LSS), astrophysics (the Sun, SN...) and terrestrial neutrino experiments (atmospheric neutrinos, reactors, short- and long-base line beams). In these Proceedings I presented in detail the analysis relative to the effects in cosmology and in supernov\ae\, focussing on the principles of the computational techniques and on the understanding of the results.


We find no evidence for a sterile neutrino in the present data; figure\fig{all} collects the {\bf presents constraints}, in the six limiting cases of mixing with a mass eigenstate $\nu_{1,2,3}$ or a flavor eigenstate $\nu_{e, \mu, \tau}$.
Every specific model of a sterile neutrino identifies a point or a region in one of these spaces or in a suitable combination of them, and should therefore be compared with the reported bounds.
In particular, I showed in Sec.\ref{LSND section} that the region of the LSND sterile neutrino is excluded by the cosmological constraints (unless they are relaxed by some non-standard modification to cosmology).

Finally, I discussed promising {\bf future probes} of sterile effects. In the context of the cosmological observations, it looks important to improve the measurements of the primordial $^4$He and Deuterium abundances, overcoming the systematic uncertainties. It looks also especially important to exploit the complementarity of the two probes, that are differently sensitive to sterile neutrinos in several regions of the parameter space, also in order to constrain the non-standard cosmological modifications. 
Moreover, future measurements of the CMB and LSS power spectra will expand the tested range of sterile parameters.
Concerning supernov\ae\, the best improvement would come from \ldots the occurring of a new explosion, which would allow to probe sterile effects through the predictions on the neutrino flux and spectrum, in regions that are not easily accessible to other tools (thanks to the long base-line and the extreme matter effects). Although progress must also come on the overall theoretical uncertainties of SN models, most of the results could be neat enough.\\
Other important probes for sterile neutrino effects are discussed in~\cite{CMSV}: among them, solar neutrino experiments at sub-MeV energies and proposed reactor and LBL experiments look most interesting.

\section*{Acknowledgments}
I thank Guido Marandella, Alessandro Strumia, Francesco Vissani, Yi-Zen Chu and the organizers of the IFAE 2004 and PASCOS'04 conferences. Work supported by the USA DoE grant DE-FG02-92ER-40704.

\end{document}